\documentclass[aps,pra,reprint,superscriptaddress]{revtex4-1}
\usepackage{amsmath}
\usepackage{hyperref}
\usepackage{physics}
\usepackage{braket}
\usepackage{color}
\usepackage[capitalise]{cleveref}
\usepackage{threeparttable}

\usepackage{graphicx}

\begin{document}

\title{Enhancement factor for the electric dipole moment of the electron in the BaOH and YbOH molecules}

\author{Malika Denis}
\email{m.denis@rug.nl}
\affiliation{Van Swinderen Institute for Particle Physics and Gravity, University of Groningen, 9747 AG, Groningen, The Netherlands}

\author{Pi A. B. Haase}
\affiliation{Van Swinderen Institute for Particle Physics and Gravity, University of Groningen, 9747 AG, Groningen, The Netherlands}

\author{Rob G. E. Timmermans}
\affiliation{Van Swinderen Institute for Particle Physics and Gravity, University of Groningen, 9747 AG, Groningen, The Netherlands}

\author{Ephraim Eliav}
\affiliation{School of Chemistry, Tel Aviv University, 69978 Tel Aviv, Israel}

\author{Nicholas R. Hutzler}
\affiliation{Division of Physics, Mathematics, and Astronomy, California Institute of Technology, Pasadena, CA 91125, USA}

\author{Anastasia Borschevsky}
\affiliation{Van Swinderen Institute for Particle Physics and Gravity, University of Groningen, 9747 AG, Groningen, The Netherlands}

\date{\today}

\begin{abstract}
Polyatomic polar molecules are promising systems for future experiments that search for violation of time-reversal and parity symmetries due to their advantageous electronic and vibrational structure, which allows laser cooling, full polarisation of the molecule, and reduction of systematic effects [I. Kozyryev and N.R. Hutzler, Phys, Rev. Lett. {\bf 119}, 133002 (2017)].  In this work we investigate the enhancement factor of the electric dipole moment of the electron ($E_\text{eff}$) in the triatomic monohydroxide molecules BaOH and YbOH within the high-accuracy relativistic coupled cluster method. The recommended $E_\text{eff}$ values of the two systems are 6.65 $\pm$ 0.15 GV/cm and 23.4 $\pm$ 1.0 GV/cm, respectively. We compare our results with similar calculations for the isoelectronic diatomic molecules BaF and YbF, which are currently used in experimental search for $P,T$-odd effects in molecules. The $E_\text{eff}$ values prove to be very close, within about 1.5 $\%$ difference
in magnitude between the diatomic and the triatomic compounds. Thus, BaOH and YbOH have a similar enhancement of the electron electric dipole moment, while benefiting from experimental advantages, and can serve as excellent candidates for next-generation experiments.
\end{abstract}

\maketitle

\section{Introduction}

The electric dipole moment of the electron ($e$EDM) is a quantity of key interest to particle physics \cite{Commins2009}. A nonzero $e$EDM, corresponding to a permanent charge separation along the electron spin axis, violates both parity, $P$, and time-reversal, $T$, symmetries. The $CPT$ theorem of quantum field theory implies that then also $CP$ invariance is broken, where $C$ is charge conjugation. Searches for the $e$EDM in low-energy experiments with atoms or molecules are therefore complementary to searches for $CP$ violation at high-energy colliders.

In the Standard Model (SM) of particle physics the $e$EDM is highly suppressed since it requires diagrams with at least four loops, in which all three generations of quarks participate \cite{Pospelov2014}. The predicted value of the $e$EDM in the SM is $d_e = \mathcal{O}(10^{-38})$ $e$~cm, which is far too small to be measured using current experimental techniques. However, extensions of the SM predict much larger values that should be within reach of state-of-the-art experiments. Hence, the discovery of a nonzero $e$EDM would serve as an unambiguous evidence for physics beyond the SM  \cite{Pospelov2014}. Another motivation for searches for the $e$EDM comes from cosmology, as additional sources of $CP$ violation, beyond the SM, are needed to explain the surplus of matter over antimatter in the Universe \cite{Sakharov1967}. 

Experiments that search for the $e$EDM have been performed for more than 50 years, with ever-increasing sensitivity. In a landmark paper \cite{Sandars1965}, Sandars pointed out that in heavy paramagnetic systems relativistic effects enhance the value of the EDM of the valence electron to a much larger EDM of the atom. Two years later, he proposed to use polar diatomic molecules, which exhibit even larger enhancement effects \cite{San67}.  One way to keep improving the sensitivity of $e$EDM experiments is to use paramagnetic systems with larger enhancement factors. For a long time, atoms, in particular thallium~\cite{Regan2002}, were at the forefront, but more recently, diatomic molecules, such as YbF \cite{Hudson2011, Kara2012}, HfF$^+$ \cite{Cairncross2017}, ThO \cite{Baron2014, Baron2017, andreev2018improved}, and BaF \cite{Aggarwal2018} are preferred, since they possess much larger enhancement factors. So far, a non-zero $e$EDM has not been detected, and the most stringent upper limit of $|d_e|<1.1 \times 10^{−29}$ $e$ cm was recently set with the ThO molecule \cite{andreev2018improved}. 
In molecules, the enhancement parameter is denoted by an effective electric field, $E_\text{eff}$, which is in essence the internal electric field seen by the valence electron. 

The statistical uncertainty of an $e$EDM experiment is typically given by 
\begin{equation}
 \sigma_d = \frac{\hbar}{e}\frac{1}{2|P|E_{\text{eff}}\,\tau \sqrt{\dot{N}T}} \ ,
\end{equation}
where $P$ is the polarization of the molecules, $\tau$ is the coherent interaction time of the molecules with the applied electric field, $\dot{N}=dN/dt$ the detection rate of the molecules, and $T$ the measurement time. The key to further progress is to increase the interaction time $\tau$ without decreasing the rate $\dot{N}$, which is possible, for example, by using beams of cold, laser-cooled diatomic molecules in $X^2\Sigma$ ground-state configuration  \cite{Shuman2010}, such as BaF \cite{Aggarwal2018}, YbF \cite{Smallman2014, Lim2018}, or RaF \cite{Isaev2010}.

Recently, it was argued \cite{Kozyryev2017} that polyatomic molecules isoelectronic to these diatomics could be the next systems of choice in the search for the $e$EDM. The reason is that polyatomic molecules generically have nearly-degenerate states of opposite parity, while maintaining large sensitivity to symmetry-violating physics (including the $e$EDM), and featuring the ability to laser cool for suitably-chosen molecules~\cite{IsaBer16,Isaev2017,Kozyryev2017,Kozyryev2018}.  The double degeneracy of the low-lying excited vibrational states, corresponding in the case of a linear triatomic to the bending of the molecule, gives rise to parity doublets with a small energy splitting, similar to the $\Omega$-doublet of the $^3\Delta_1$ metastable states of the HfF$^+$~\cite{Cairncross2017} and ThO~\cite{andreev2018improved} molecules used presently in $e$EDM measurements (Fig. 1).   This particular structure allows for full polarisation of the molecule in comparatively low electric fields, and
 for the existence of internal comagnetometer states \cite{MeyBohDes06,LeeMeyPau09,EckHamKir13,andreev2018improved,Cairncross2017}.  In diatomics this relies on electronic structure that is not amenable to laser cooling for $e$EDM-sensitive species, yet generically arises in polyatomic molecules regardless of electronic structure.  Note that in symmetric top molecules, such as YbOCH$_3$ or BaOCH$_3$, this splitting arises even in non-vibrating states due to rigid-body rotations about the symmetry axis, resulting in even smaller splittings~\cite{Kozyryev2017}. 
 
 \begin{figure}
	\centering
		\includegraphics[width=0.3\textwidth]{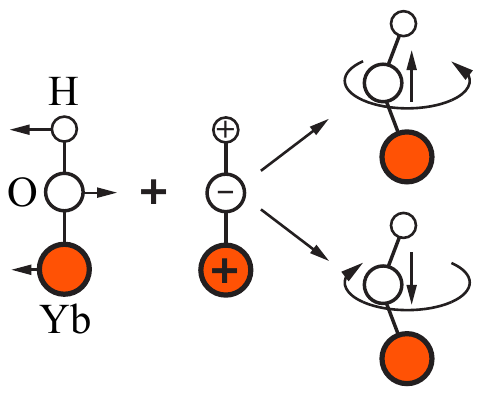}
	\caption{The physical eigenstates of the doubly-degenerate bending mode in a linear triatomic molecule possess angular momentum with non-zero projection on the internuclear axis.  This angular momentum interacts with the molecular rotation and gives rise to a parity doublet split by $\sim$20~MHz.}
	\label{fig:PolyatomicAngularMomentum}
\end{figure}
 
 The measurement of the energy shift in the two parity states mimics the reversal of the applied electric field experienced by the electron, thereby removing the need to reverse any external fields, and avoiding systematic errors. The advantage of the polyatomic molecules over the $^3\Delta_1$ diatomics is that these systems are amendable to laser-cooling, as was proposed theoretically for RaOH \cite{Isaev2017}, YbOH \cite{Kozyryev2017}, and a number of other molecules \cite{IsaBer16,Kozyryev2016Poly}, and demonstrated experimentally for SrOH \cite{KozBauMat17}. The sensitivity of these systems to the $e$EDM  (expressed in the magnitude of $E_\text{eff}$) is expected to be very similar to that in the corresponding isoelectronic diatomics, based on the assumption that the main contribution originates from the non-bonded electron localised on the heavy atom and the ligand is expected to play a minor role \cite{IsaBer16, Isaev2017, Kozyryev2017}.

In this work, we investigate the barium and ytterbium monohydroxide molecules, BaOH and YbOH, and calculate their $e$EDM enhancement factors $E_\text{eff}$ within the finite-field method in the framework of the relativistic coupled cluster approach. Particular attention is paid to the evaluation of the uncertainty of the computations. In addition, we compare our results for BaOH and YbOH with the values obtained with the same method for the corresponding diatomic molecules, BaF and YbF, which possess the same $X^2\Sigma$ ground-state configuration. The aim of this comparison is to test the assumption that the values of $E_\text{eff}$ are very similar in these isoelectronic systems.

The $E_\text{eff}$ values of BaOH and YbOH were recently calculated using the combination of the  zeroth-order regular approximation (ZORA) and either the complex generalised Hartree-Fock (cGHF) or the Kohn-Sham method (cGKS) \cite{GauBer18}. The authors estimate the accuracy of the approach as about 20$\%$. 

\section{Methodology}
The value of the $e$EDM is extracted from the experimental energy shift $\Delta E$  through 
\begin{equation}
 d_e = \Delta E / E_{\text{eff}} \ ,
\end{equation}
where $E_\text{eff}$ is the system-dependent relativistic enhancement factor that describes the interaction of the EDM of the unpaired electron with the molecular internal electric field. $E_{\text{eff}}$ can not be measured directly and is obtained from electronic structure calculations.

Following Lindroth's stratagem II \cite{Lindroth1989}, the $e$EDM Hamiltonian written as a one-particle operator for electron $i$ reads
\begin{equation}
 H^\text{EDM} = 2{c} d_e \sum_{i=1}^n\imath\gamma^0(i)\gamma^5(i)\mathbf{p}^2(i),
\end{equation}
where $\gamma^{0,5}$ are the standard Dirac matrices and $\mathbf{p}(i)$ is the momentum operator for electron $i$.

The $E_\text{eff}$ calculations are performed in the framework of the finite field approach \cite{Coh65,PopMcIOst68,Mon77}, recently extended to the $E_{\text{eff}}$ property \cite{AbePraDas18,Haa18}.  
In this approach we replace the electron EDM $d_e$ in $H^\text{EDM}$ by a perturbation parameter $\lambda$, and add it to the molecular Hamiltonian $H^{(0)}$,


\begin{equation}
H = H^{(0)}+\lambda \cdot 2{c} \sum_{i=1}^n\imath\gamma^0(i)\gamma^5(i)\mathbf{p}^2(i) 
\end{equation}
$H^{(0)}$ is the relativistic Dirac-Coulomb (DC) Hamiltonian:
\begin{equation}
 \hat{H}^{(0)}=\sum_i^n\left[c\boldsymbol{\alpha}_i\cdot\textbf{p}_i+\beta_i c^2+V_{\text{nuc}}(\textbf{r}_i)\right]
 +\sum_{i<j}\frac{1}{r_{ij}},
 \end{equation} 
where $n$ is the number of electrons, $\boldsymbol{\alpha}_i$ and $\beta_i$ are the Dirac matrices, and $V_{\text{nuc}}$ the Coulomb potential, which takes into account the finite size of the nuclei, modelled by Gaussian charge distributions \cite{VisDya97}.





Provided that the values of the perturbation parameter $\lambda$ remain sufficiently small to guarantee linear behaviour of the energy the effective electric field can be obtained numerically, according to the Hellmann-Feynman theorem, from the first derivative of the energy with respect to $\lambda$, {\em viz.}
\begin{equation}
\label{eeff_ff}
 E_{\text{eff}} \equiv \bra{\Psi}  H^\text{EDM'} \ket{\Psi}=\left.\frac{dE(\lambda)}{d \lambda}\right|_{\lambda=0} \ .
\end{equation}


 
 

\section{Computational details}
All the calculations were performed with a modified version of the DIRAC17 program package \cite{DIRAC17}. Experimental atomic distances were employed for 
both molecules which are linear in their $X^2\Sigma$ ground-state, d$_\text{Ba-O}=2.201$\AA{}, d$_\text{O-H}=0.923$\AA{} for BaOH 
\cite{KinseyNielsen1986}  
and d$_\text{Yb-O}=2.0369$\AA{}, d$_\text{O-H}=0.9511$\AA{} for YbOH \cite{Brutti2005,Nakhate2019}. A new measurement of rotational spectrum of YbOH was published after the present calculations were completed \cite{NakStePil18}. The reported there  d$_\text{Yb-O}=2.0397$\AA{} is very close to the value that we used in our study.

The calculations were performed within the finite-field approach outlined above. For this purpose, the same calculation was repeated three times, with applied fields of $-10^{-9}$, $0$, and $1.10^{-9}$ a.u., chosen to ensure that the linear behaviour of the total energy with respect to this perturbation (\cref{eeff_ff}) applies. To support such small field strengths, the convergence criterion of the coupled cluster amplitudes was set to $10^{-12}$. The effective electric field was then determined by linear fitting of the three obtained energies.

We used two variants of the relativistic coupled cluster approach: the standard single reference coupled cluster with single, double, and perturbative triple  excitations (CCSD(T)) \cite{VisLeeDya96,Visscher1998} and the multireference Fock Space Coupled Cluster (FSCC) \cite{Visscher2001}.
We compare these two approaches, along with second-order many-body perturbation theory  (M\o{}ller-Plesset theory, MP2) \cite{MolPle34} results and the uncorrelated Dirac-Hartree-Fock (DHF) values neglecting spin polarization. 
In the CC and MP2 calculations, if not stated otherwise, all electrons were correlated and the virtual space cutoff was set to 2000 a.u. 

We have investigated the effect of the basis set on the calculated $E_{\text{eff}}$ by employing Dyall's relativistic uncontracted valence basis sets of varying quality (vNz  with N$\in\{2,3,4\}$) \cite{Dyall2009,GomDyaVis10,Dyall2016} and augmenting them manually in certain cases.



\section{Results}

In accordance with the assumption in Ref.~\cite{Kozyryev2017}, we expected BaOH to be similar to BaF (and YbOH to YbF) and their effective electric fields to exhibit the same behaviour with respect to the variation of the different computational parameters. Our choice of the computational scheme was thus motivated by the extensive studies performed on BaF in the context of \textit{P}-- and \textit{P,T}--odd interactions \cite{HaoIliEli18,Haa18}. In particular, we correlated all the electrons and set the virtual space cutoff to 2000 a.u due to our finding that both core electrons and high virtual orbitals provide a significant contribution to the calculated interaction coefficients when aiming for highest possible accuracy. Here we perform further investigations on BaOH and YbOH themselves in order to set the uncertainty on their predicted value of $E_\text{eff}$.

 
Below we address each computational parameter separately (subsections IV.A--IV.D) and provide the recommended $E_\text{eff}$ values and uncertainty estimates in subsection IV.E.

\subsection{Basis set size}

Table I contains the calculated $E_\text{eff}$ values of BaOH and YbOH for varying quality basis sets. For both systems, we observe an increase of a few percent in the $E_\text{eff}$ value when going from v2z to v3z basis set. Switching to the v4z basis causes a decrease of around 1 \% in the $E_\text{eff}$ of BaOH; such a non-monotonic behaviour with respect to the basis set size was also observed for the $E_\text{eff}$ of BaF \cite{Haa18}. We thus do not attempt extrapolation to the complete basis set limit (CBSL) in this case. In YbOH the v4z value continues the trend and is higher by about 1 \% than the v3z one. Therefore, extrapolation of the obtained results to CBSL is appropriate in this case. We used the exponential extrapolation scheme for the DHF part of the energy \cite{HalHelJor99} and the inverse cubic extrapolation for the correlation energies \cite{HelKloKoc97}. The extrapolated $E_\text{eff}$ of 23.91 GeV/cm is very close to the v4z value, indicating saturation with respect to the basis set. 

\begin{table}[h!]
 \caption{Calculated $E_\text{eff}$ [GV/cm] of BaOH (CCSD(T)) and YbOH (FSCC) within basis sets of varying quiality and the X2C and the 4-component Hamiltonians. All the electrons were correlated and a virtual cutoff of 2000 a.u. was used}
 \label{x2cBaOH}
 \begin{tabular}{llcc}
 \hline 
 Basis set& Hamiltonian &$E_\text{eff}$(BaOH)& $E_\text{eff}$(YbOH)\\
 \hline
 v2z&x2c&$6.393$&$22.68$\\
      &4c&$6.421$&$22.68$\\
 v3z&x2c&$6.533$&$23.50$\\	
      &4c&$6.552$&$23.54$\\
v4z&x2c&$6.468$&$23.69$\\	
      &4c&$6.496$&$23.76$\\ 
 opt-v4z&x2c&$6.524$&-\\	
      &4c&$6.533$&-\\ 
s-aug-v4z&4c&$6.496$&-\\
CBSL&x2c&-&23.84\\	
      &4c&-&23.91\\    
            &$\Delta$ Gaunt&-0.112&$-$0.35\\
\hline
 \end{tabular}	
\end{table}

It is known that a high quality description of the electronic wave function in the nuclear region is essential for obtaining reliable results for parity-violating effects \cite{LaeSch99,BorIliDzu12-2,BorIliDzu13}, and thus large exponent (tight) functions may play an important role. In the tests that we carried out on the BaF molecule we found that indeed addition of a single large-exponent $f$ function provides a contribution of about 1 $\%$ to both the $W_A$ and  $E_\text{eff}$ parameters \cite{HaoIliEli18,Haa18}. We thus add such a function to the v4z basis set and designate the obtained basis set as opt-v4z; the augmentation raises the $E_\text{eff}$ in BaOH by $0.6\%$. This optimised basis set will be used for the determination of the final value of the effective electric field in the $X^2\Sigma$ state of BaOH. To get an estimate of the effect of adding further tight $f$ functions, or tight functions of other symmetries, we also performed a calculation with cv4z (core-valence 4z basis), which contains additional 3 tight $f$, 2 tight $g$, and 1 tight $h$ functions for Ba, and 2 tight $d$ and 1 tight $f$ function for oxygen (saturation with respect to the tight $s$, $p$, and $d$ functions was tested for $E_\text{eff}$ of BaF in Ref. \cite{Haa18}). The cumulative effect of all these additional tight functions is 0.03 GV/cm, or 0.5 \%.  

In case of Yb (and the rest of the lanthanides) the optimised dyall-vNz basis sets are quite dense  and notably include an extended set of $f$-type functions  in order to obtain a good description of the $4f$ shell which is close in energy to the valence shells \cite{GomDyaVis10}. We estimate the effect of including further tight functions by performing a calculation with ae4z basis set (all-electron 4z basis containing additional 1 tight f, 3 tight g and 1 tight h functions for Yb and 2 tight d and 1 tight f for  O). Due to significant computational expense this investigation was carried out for YbF instead of YbOH.
In this case, augmentation has negligible effect (less than 0.05 $\%$). 
 
The final parameter that is tested here is the effect of adding diffuse functions, which are important for the description of the region far from the nucleus. We have tested the s-aug-v4z basis sets for both molecules (s-aug stands for adding a single diffuse function for each symmetry in the basis set). For both systems the effect was minuscule (less than 0.1 \%), which leads us to conclude saturation with respect to the diffuse functions.




\subsection{Treatment of relativity}

In order to reduce the computational cost of the coupled cluster calculations, one could be willing to replace the relativistic four-component Hamiltonian with a high quality approximation, such as the infinite order exact two-component (X2C) Hamiltonian \cite{IliSau07,Sau11}. This Hamiltonian is routinely employed in conjunction with the atomic mean field integral (AMFI) code, which provides the two-electron spin-orbit contributions \cite{IliKelVis01}. The X2C approach allows a  speed-up in the calculations
while reproducing very well the results obtained using the 4-component DC Hamiltonian \cite{Bast2009,SikVisSau09,Knecht2010}.

Here we compare the X2C results to the full 4-component  $E_\text{eff}$ values in BaOH and YbOH (~\cref{x2cBaOH}) for varying quality basis sets. We find that the difference between the two methods is $0.2-0.4 \%$, which supports the use of the X2C approach for this property. This can be useful in particular in case of large molecules, where the system size makes the expense of 4-component calculations prohibitive. We, however, continue to employ the 4-component Hamiltonian throughout the rest of this study, as it is still feasible for the systems that we are interested in.

In addition we test the influence of the Gaunt term  \cite{Gau29} on our results. 
This term is the analytic frequency independent approximation to the single transverse photon  exchange between electrons in the Feynman gauge. The same  approximation in the Coulomb gauge
(is known as Breit interaction) corrects the 2-electron part of the Dirac-Coulomb Hamiltonian up to order $(Z \alpha)^2$ \cite{Bre29}. 
Although this does not hold for the
Feynman gauge expression even if the frequency dependence is included \cite{sucher1988}, still the Gaunt term, which is technically much more easily implementable in computer codes,  is regarded to be a good approximation to the full QED based single transverse photon  exchange interaction between electrons.
The Breit term is not yet implemented in the DIRAC program, and the Gaunt interaction is included self-consistently at the DHF step. This contribution reduces the effective electric field by about 0.1 GV/cm for BaOH and 0.35 GV/cm for YbOH (~\cref{x2cBaOH}); we include this contribution in the final recommended $E_\text{eff}$ values.


\subsection{Virtual space cutoff}

In standard coupled cluster applications the virtual space cutoff is usually set at around ~30 a.u., at which point saturation of most properties with respect to the correlation space size is reached (in particular when core electrons are excluded from the correlation procedure). However, it was found that inclusion of the high lying virtual orbitals is important for the correlation of the core electrons for parity violating properties \cite{SkrMaiMos17,HaoIliEli18}. 
We thus set the cutoff of the virtual space to 2000 a.u. in all the calculations and also investigated the ensuing error with respect to a nontruncated virtual space. Results of the study made at the triple-zeta level are shown in \cref{baohcutoff} for the two systems. For BaOH the inclusion of twenty more spinors (corresponding to a virtual space cutoff of 6000 a.u.) increases slightly the value of $E_\text{eff}$ by $0.024$ or less than $0.4\%$ in magnitude. Further increase of the virtual cutoff to 10000 a.u., \textit{i.e.}, adding eighteen more spinors, increases the magnitude of $E_\text{eff}$ by less than $0.15\%$. Thus, we can consider the results converged at the 10000 a.u. level and stop the study at this point. 

Globally, $E_\text{eff}$ at a cutoff of 10000 a.u. is $0.55\%$ higher with respect to the value obtained with the 2000 a.u. virtual cutoff employed in the rest of this work. We expect similar correction at the optimized-v4z level, and thus we will consider this figure as the uncertainty on the final value due to the cutoff set on the virtual spinor space.

\begin{table}[h!]
 \caption{
 Calculated $E_\text{eff}$ of BaOH (CCSD(T)) and YbOH (FSCC) using various cutoffs of the virtual space. All the electrons were correlated and dyall-v3z basis sets were used.}
 \label{baohcutoff}
 \begin{tabular}{llclr}
 \hline
 &{Cutoff} &   $\#$ spinors	& Added spinors	&$E_\text{eff}$ [GV/cm]\\
 \hline
 \textbf{BaOH~} 
&$2000$ a.u. &  $381$	&	&$6.552$\\
&$6000$ a.u. &  $401$	& + Ba $d,s,p$; O $s$	&$6.576$\\
&$10000$ a.u.	& $419$ & + Ba $d,s,p$	&$6.585 $\\ 
\cline{2-5}
 \textbf{YbOH}
& $2000a.u.$ &  $568$		& 			&$23.58$\\
&$6000a.u.$ &  $590$		& +Yb $s,p,d,s$; O $s$ &$23.61$\\
 \hline
 \end{tabular}	
\end{table}

A similar test was performed for YbOH; increasing the cutoff from 2000 a.u. to 6000 a.u. entailed only a $0.12\%$ increase in the value of $E_\text{eff}$ and thus we did not increase the correlation space further. 

 
\subsection{Treatment of electron correlation}

\cref{baohcorrel} contains the $E_\text{eff}$ of BaOH, calculated on different levels of correlation treatment (these results were obtained with the optimised basis set). Neglect of correlation underestimates the $E_\text{eff}$ by about 25 \%, compared to CCSD, and MP2 reduces this error to 10 $\%$.
In the single reference coupled cluster approach, as implemented in the DIRAC program package, triple excitations are addressed in a perturbative fashion on three levels that differ by the terms included in the perturbation theory. The standard scheme (CCSD(T), \cite{RagTruPop89}) includes all fourth-order terms and part of the fifth-order terms while CCSD-T \cite{DeeKno94} includes further fifth-order terms and CCSD+T includes fourth-order terms only \cite{UrbNogBar85}. Results displayed in \cref{baohcorrel} show that inclusion of triple excitations lowers the $E_\text{eff}$ by up to  1.7 \% compared to CCSD, but the choice of the scheme for treatment of triple excitations does not have a strong influence on the results since the three values are within $0.3\%$ of each other. We will, however, use the spread in these values for estimating the uncertainty of the recommended $E_\text{eff}$.  

BaOH has a single valence electron in the $\sigma$ orbital and thus two different FSCC computational schemes are appropriate for this system. The first one is designated FSCC(0,1), where the calculation begins with BaOH$^+$, and an electron is added in the coupled cluster procedure to obtain the correlated energy of the neutral system. This extra electron can be added to the lowest $\sigma$ orbital, corresponding the the minimal model space (\textit{Min}) or it can also be allowed to occupy some higher states, thus yielding a number of energy levels and also improving the description of the ground state energy and properties. In order to test this effect on $E_\text{eff}$ we have also used a larger model space, \textit{Ext}, which included 2 $\sigma$, 2 $\pi$, and 2 $\delta$ spinors (i.e. the 12 lowest spinors). The sector (0,1) results are very close to the CCSD(T) values; superior performance of FSCC in particle sectors compared to single reference CCSD has been observed in the past \cite{EliBorKal17}, and also in the context of parity violating properties \cite{HaoIliEli18}.  The results obtained using two different model spaces were within 0.3 \% of each other indicating that the minimal model space is sufficient here. 

The second computational scheme suitable for BaOH is the (1,0) sector of FSCC, where the calculation commences with negatively charged BaOH$^{-}$, and an electron is removed in the coupled cluster procedure. In this case the obtained $E_\text{eff}$ is actually higher than that for the CCSD approach, rather than lower (or similar), as one would expect. The reason for the comparatively poor performance of this approach is probably that the basis sets used in this work are  inadequate for a good description of a negatively charged reference state ion without further augmentation by diffuse functions. We thus ignore this approach in the uncertainty estimation for BaOH.




\begin{table}[h!]
 \caption{ Calculated $E_\text{eff}$ of BaOH using various correlation methods. All the electrons were correlated in the CC procedure, the virtual space cutoff was set to 2000 a.u. and the opt-v4z basis set was used. The final recommended value is given in bold font.
}
 \label{baohcorrel}
 \begin{tabular}{llc}
 \hline
 Method&&$E_\text{eff}$ [GV/cm]\\
 \hline
        &DHF         &$4.836$\\
               &MP2        &$5.977$\\
    Single reference&CCSD		&$6.627$ \\
    				&CCSD(T)	&$6.533$\\
    				&\textbf{CCSD(T)+Gaunt}	&\textbf{6.421} \\
				&CCSD+T		&$6.515$ \\
				&CCSD-T 	&$6.538$ \\
	Multireference		&FSCC(0,1) \textit{Min}	&$6.538$ \\
		&FSCC(0,1) \textit{Ext}	&$6.515$  \\
				&FSCC(1,0) 	&$6.716$\\	
\hline
\end{tabular}
\end{table}


The multi-reference character of the ground-state of YbOH (and YbF) makes the perturbative treatment of triple excitations  highly unstable and yields unphysically large and unreliable results, preventing the use of  the standard CCSD(T) approach, and necessitating the switch to the multireference Fock space coupled cluster. The instability of the CCSD(T) method for ytterbium compounds was also observed in earlier calculations of the electric field gradients in YbF \cite{Pasteka2016}.

 Our analysis of the coupled cluster results for YbOH is thus restricted to the FSCC and CCSD values. Here we also examine how the basis set quality affects the results obtained with different correlation approaches. The results are collected in \cref{ybohcorrel}. The DHF and MP2 values are practically insensitive to the basis set size, and we thus only show these values for the v2z quality basis. Like in case of BaOH, the DHF value is lower by about 25 \% than the CCSD result, while this error is ~5 \% for MP2. 
 
  The trend in the values of $E_\text{eff}$ calculated with single reference CCSD  is highly irregular with respect to the basis set, with a decrease of 15 \% when moving from v2z to v3z, and an increase of ~20 \% when switching to v4z. We find very similar behaviour in YbF \cite{Haa18}, as did Abe et al. in Ref. \cite{AbeGopHad14}. This instability, in particular for the v3z basis, can be attributed to the multireference character of the ground state configurations of these systems, which we were able to diagnose via the large $T1$ values. $T1$ diagnostic is the norm of the vector of $T_1$ (single excitation) amplitudes, scaled to be independent of the number of correlated electrons N. This value can be used for estimating the reliability of results obtained from a single-reference-based electron correlation procedure, such as CCSD. It was found empirically that a large $T1$ ($T1>$ 0.02) indicated the need to employ a multireference  approach \cite{LeeTay89}. In the YbF and YbOH compounds we found $T_1=0.1$ for the v3z basis set. The multireference character of these systems is due to a low-lying excited state stemming from the $f^{13} s^2$ configuration that interacts strongly with the ground $X ^2 \Sigma ^+$ $f^{14} s^1$ state \cite{DolStoHei92,GomDyaVis10}.  Therefore we use the multireference FSCC, which is more appropriate for this system, both in our investigations and for the final recommended value. The FSCC values change smoothly with the basis set improvement, as discussed above (Section IV.B). We find that here, like in BaOH, the model space size has negligible (0.25 \%) influence on the results. The FSCC (1,0) result is about 4 $\%$ higher than the (0,1) value. 
 

\begin{table}[h!]
 \caption{Calculated $E_\text{eff}$ of YbOH using various correlation methods and basis sets. All the electrons were correlated in the CC procedure and the virtual space cutoff was set to 2000 a.u. The final recommended value is given in bold font.}
\label{ybohcorrel}
\begin{tabular}{lllc}
 \hline
 Basis &&Method &$E_\text{eff}$ [GV/cm]\\
 \hline
  v2z&                  &DHF        &$18.01$\\
    &                   &MP2        &$21.30$\\
    &	&CCSD		&$22.54$\\
	&	&FSCC(0,1) \textit{Min}	&$22.68$ \\
   \cline{2-3}
v3z   &	&CCSD		&$19.24$ \\
	&	&FSCC(0,1) \textit{Min}	&$23.54$ \\
\cline{2-3}
v4z    & &CCSD		&{$23.73$} \\
	& &FSCC(0,1) \textit{Min}	&$23.76$ \\
    &					&FSCC(0,1) \textit{Ext}	&$23.70$ \\
    &					&FSCC(1,0)  &$24.62$\\
    CBSL& 		&\textbf{FSCC(0,1) \textit{Ext}+Gaunt} 	&23.35 \\
\hline
\end{tabular}	
\end{table}

\subsection{Recommended values and error estimation}

We will consider the 4c-CCSD(T) result obtained with the opt-v4z basis set and a virtual cut-off of 2000 a.u., corrected for the Gaunt contribution, as our final recommended value of the effective electric field of BaOH (shown in bold font in \cref{baohcorrel}). The choice of CCSD(T) over the other schemes for treatment of triple excitations is motivated mainly by the popularity of this approach, as the numerical difference between the schemes is close to negligible.

In case of YbOH we take the 4c-FSCC result obtained with the extended model space, 2000 a.u. virtual space cut-off, extrapolated to the basis set limit and corrected for Gaunt contribution as the final value (\cref{ybohcorrel}).

A challenging step of this work is the evaluation of the uncertainty of the calculated enhancement factors $E_\text{eff}$, which is important for the interpretation of the experimental measurement of the $e$EDM. Since $E_\text{eff}$ can only be determined by theoretical calculations, we can not use comparison to experiment to evaluate the performance of the employed approach. Therefore, we need to devise an alternative method to assign an uncertainty on the predicted values; one possibility is to base this uncertainty on computational considerations. In this work we tested the influence of various parameters of the procedure employed; these tests allow us to assign an error due to each of these parameters. \\

\textit{Basis set quality} For both systems the results seem to saturate with respect to the basis set quality, with the difference between v3z and v4z basis on the order of a single percent. For BaOH we take that value to estimate possible effect of going to the v5z basis and further. In case of YbOH we have performed extrapolation to the complete basis set limit, and hence we take the difference between the CBSL result and the v4z value as the error estimate. We also estimate the effect of basis set augmentations with diffuse and tight functions, by taking the difference between the results obtained for the present basis sets and those augmented further. For BaOH, the uncertainty due to neglect of further large-exponent functions is 0.033 GV/cm (Section IV.A), and that caused by the lack of further diffuse functions is 0.001 GV/cm; the effect on YbOH is even smaller.

\bigskip

\textit{Virtual space cut-off} Here, we take twice the difference between the values calculated with cut-offs of 2000 a.u. and 6000 a.u. as the uncertainty.\\

\bigskip

\textit{Full triples and higher-order excitations} In case of BaOH, we are able to estimate the effect of triple excitations by including these perturbatively, via CCSD(T) approach. We take twice the difference between CCSD+T and CCSD-T ($0.05$ GV/cm, which is also about half of the total perturbatve triples contribution) as the uncertainty due to incomplete treatment of triple 
and higher 
excitations.

For YbOH, perturbative calculation of triple excitations has proved to be intractable. Therefore, to obtain an estimate of the effect of neglecting triple and higher excitations we take the difference between the FSCC sector (0,1) and sector (1,0) results. 
Use of this procedure is motivated by the fact that the results for the two schemes should be equal to the full CI value (and hence to each other) in the case when all possible multielectronic dynamic excitations are being accounted for \cite{stolarczyk1985coupled}, and thus the difference between the two values indicates the weight of the neglected excitations.
At 4 \%, this is a single largest source of error found in this work; however, the complex nature of the ground state of YbOH justifies the conservative uncertainty evaluation.

\bigskip

\textit{Treatment of relativity} We assume that the effect of replacing the Breit term by the Gaunt interaction and neglecting QED effects is not more than the contribution of the Gaunt term itself 
(0.112 GV/cm for BaOH and 0.35 for YbOH).\\




\begin{table}[h!]
 \caption{Summary of the most significant error sources in $E_\text{eff}$ in BaOH and YbOH (GV/cm) (see text for details).
}
 \label{baoherror}
 \begin{tabular}{lllcc}
 \hline
\multicolumn{2}{l}{Error source}	    &		& ~BaOH~ 			&	  ~YbOH \\
\hline
\multicolumn{2}{l}{Basis quality}    &				&0.056		&	0.13	\\
\multicolumn{2}{l}{Basis augmentations}	&			& 		&	\\
&Tight functions	    	&			& 0.033		& 0.01\\
&Diffuse functions	    	&			& 0.006		&	0.02\\
\multicolumn{2}{l}{Correlation}  & & & \\
&Virtual space cut-off	    &	&0.048		&   	0.06	\\
&Residual triples and&	&		&	\\
&higher excitations	    &		&	0.055& 0.92		\\
\multicolumn{2}{l}{Relativity} &  &0.112&0.35\\ 

\cline{4-5}
\textbf{Total~~~}&	    &		&			\textbf{0.150}	& \textbf{0.98}		\\
\hline
 \end{tabular}
\end{table}

Combining the above sources of error listed in Table V (and assuming them to be independent), the total uncertainty for BaOH is 0.150 GV/cm and for YbOH 0.98 GV/cm, corresponding to 2.2 \% and 4.2 \%, respectively. Thus, the final recommended $E_\text{eff}$ values of the two molecules are 6.42 $\pm$ 0.15 GV/cm and  23.4 $\pm$ 1.0 GV/cm.

\begin{table}[t!]
\caption{$E_\text{eff}$ (GV/cm) values of BaOH and YbF and of their isoelectronic diatomic analogues, BaF and YbF. For triatomics, final recommended values are shown; the calculations for the diatomics were performed within the same scheme (see text for details).}
\label{comparison}
  \begin{tabular}{llllll}
  \hline
  \hline
  BaOH&CCSD(T)&6.421&YbOH&FSCC&{23.35}\\
  BaF$^*$&CCSD(T)&6.332&YbF$^*$&FSCC&23.56\\
  \hline
  \hline
  \end{tabular}
  \begin{flushleft}
  $^*$Ref. \cite{Haa18}
  \end{flushleft}
\end{table}

\subsection{Comparison with diatomics}

In \cref{comparison} our final values of the effective electric field of BaOH and YbOH are shown, together with the corresponding values for their analogous diatomic molecules, BaF and YbF \cite{Haa18}, calculated within the same method and using the same computational parameters. We conclude that changing fluorides for monohydroxides does not affect the magnitude of $E_\text{eff}$ by more than $1.5\%$, due to the fact that the main contribution to this property stems from the unpaired electron localised on heavy nucleus, as was indeed suggested in earlier works \cite{Kozyryev2017, Isaev2017}. Furthermore, the F and OH groups are similar in charge and size, and thus polarize the Ba orbitals in the same fashion. This implies that the enhancement factors remain sufficiently large for BaOH and YbOH to be suitable for the search of the $e$EDM.  

The present results can be compared to the recent study of Gaul and Berger \cite{GauBer18}. For BaOH the cGHF (6.9 GV/cm) and cGKS (6.2 GV/cm) values of Ref. \cite{GauBer18} are very close to each other, and, surprisingly, also close to our CCSD(T) result, rather than to DHF. In case of YbF there is a large difference between the cGHF and the cGKS $E_\text{eff}$ (23.6 and 17.7 GV/cm, respectively), stemming from the multireference character of this system; the cGHF result is again close to the present FSCC prediction. Ref. \cite{GauBer18} also finds that the values of $E_\text{eff}$ for the triatomic compounds are very close to those of their diatomic analogues.

\section{Conclusions}

In this work, we performed high-accuracy relativistic coupled cluster calculations of the effective electric field in the triatomic BaOH and YbOH molecules. The effect of various computational parameters on the obtained result was explored. Noteworthy is the finding that while the single reference CCSD(T) approach performs well for BaOH, the multireference character of the ground state of YbOH made it necessary to use the Fock space coupled cluster method, which is more suitable for multireference systems. The other investigated parameters include the basis set quality,  the size of the active space in the CC treatment, the performance of X2C vs. the full 4-component Dirac Hamiltonian, and the effect of the Gaunt term. These investigations allowed us to set an uncertainty on the calculated values, and the final recommended $E_\text{eff}$ are 6.42 $\pm$ 0.15 GV/cm and  23.4 $\pm$ 1.0 GV/cm for BaOH and YbOH, respectively.


We find that the magnitude of the effective electric field in these systems is very close to that in their corresponding isoelectronic diatomic systems BaF and YbF, while their laser coolability and the structure of their first vibrational excited states bring important additional experimental assets that may well make them better candidates for the search for the $e$EDM and other $CP$-violating phenomena.  This gives further evidence that polyatomic molecules are indeed promising systems for next-generation searches for fundamental symmetry violations.


%

\end{document}